\documentclass[11pt]{article}
\usepackage{vietnam}
\usepackage{amssymb}
\usepackage{url}

\usepackage[flushleft]{threeparttable}

\usepackage{amsmath,amsfonts}
\usepackage{txfonts}
\usepackage{graphicx}

\def\Journal#1#2#3#4{{#1} {\bf #2}, #3 (#4)}

\def\be{\begin{equation}}
\def\ee{\end{equation}}
\def\bea{\begin{eqnarray}}
\def\eea{\end{eqnarray}}

\newcommand{\Photo}{\includegraphics[height=35mm]{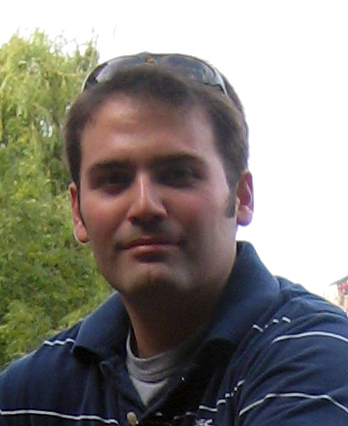}}

\begin{document}
\vspace*{4cm}
%\title{Planck 2013 results. XXVIII. The Planck Catalogue of Compact Sources}
\title{The \textit{Planck} Catalogue of Compact Sources}

\author{\textit{Planck} Collaboration, presented by M.~L\'opez-Caniego}

\address{Instituto de F\'isica de Cantabria, CSIC-UC, Avda. los Castros, s/n, \\
E-39005 Santander, Spain}

\maketitle\abstracts{ The \textit{Planck} Catalogue of Compact Sources (PCCS) is the catalogue of sources detected in the first 15 months of \textit{Planck} operations, the ``nominal'' mission. It consists of nine single-frequency catalogues of compact sources, both Galactic and extragalactic, detected over the entire sky. The PCCS covers the frequency range 30 -- 857\,GHz with higher sensitivity and better angular resolution  than previous all-sky surveys in the microwave band.  It is 90\% complete at 180 mJy in the best channel, and the resolution ranges from 32.88 to 4.33 arc minutes.  By construction its reliability is $>80\%$, and more than 65\% of the sources have been detected at least in two contiguous \textit{Planck} channels. Many of the \textit{Planck} PCCS sources can be associated with stars with dust shells, stellar cores, radio galaxies, blazars, infrared luminous galaxies and Galactic interstellar medium features. Here we summarize the construction and validation of the PCCS, its contents and its statistical characterization.}

\def\quijote{{\it QUIJOTE\/}}
\def\Planck{{\it Planck\/}}
\def\eps{\varepsilon}
\def\aap{A\&A}
\def\apj{ApJ}
\def\apjs{ApJS}
\def\apjl{ApJL}
\def\apss{Ap\&SS}
\def\mnras{MNRAS}
\def\aj{AJ}
\def\nat{Nature}
\def\aaps{A\&A Supp.}
\def\prd{Phys. Rev. D}

%\keywords{cosmic microwave background, polarisation, cosmological parameters,
 % early Universe, telescope, instrumentation }

%%%%%%%%%%%%%%%%%%%%%%%%%%%%%%%%%%%%%%%%%%%%%%%%%%%%%%%%%%%%%
\section{Introduction}
\label{sec:intro}
The \textit{Planck} Catalogue of Compact Sources (PCCS), Planck Collaboration Results 2013 XXVIII~\cite{pccs}, comprises nine single-frequency source lists, one for each \textit{Planck}\, frequency band, Planck Collaboration Results 2013 I~\cite{planckmain}. It contains high reliability sources, both Galactic and extragalactic, detected over the entire sky. The PCCS differs in philosophy from the Early Release Compact Source Catalogue (ERCSC), Planck Intermediate Results VII~\cite{planck2012-VII},  in that it puts more emphasis on the completeness of the catalogue, without greatly reducing the reliability of the detected sources ($> 80\%$ by construction).

The data obtained from the \textit{Planck} nominal mission between 2009 August 12 and 2010 November 27 have been processed into full-sky maps by the Low Frequency Instrument (LFI; 30--70\,GHz) and High Frequency Instrument (HFI; 100--857\,GHz) Data Processing Centres (DPCs). The data consist of two complete sky surveys and 60\% of the third survey. This implies that the flux densities of sources obtained from the nominal mission maps are the averages of at least two observations. The nine \textit{Planck} frequency channel maps were used as input to the source detection pipelines. For the highest-frequency channels (353, 545 and 857\,GHz), a model of the zodiacal emission (Planck Collaboration XIV 2013~\cite{plxiv}) was subtracted from the maps before detecting the sources.

Compact sources were detected in each frequency map by looking for peaks after convolving with an optimal Mexican Hat Wavelet 2 (MHW2) filter (M. L\'opez-Caniego {\it et al.}~\cite{comp}) that preserves the amplitude of the source while reducing the large-scale structure (e.g., diffuse Galactic emission) and small-scale fluctuations (e.g., instrumental noise) in the vicinity of the sources.  The MHW2 is a robust filter and performs well at all Galactic latitudes. It has only one free parameter, the width of the filter $R$, that is optimized locally, and is less sensitive than other filters to artefacts (e.g., missing pixels) or very bright structures in the image, like those found in the Galactic plane.

\section{Selection criteria}
The source selection for the PCCS is made on the basis of the signal-to-noise ratio (S/N). However, the background properties of the \textit{Planck} maps vary substantially depending on frequency and sky area. Up to 217\,GHz, the CMB is the dominant source of confusion at high Galactic latitudes. At high frequencies, confusion from Galactic foregrounds dominates the noise budget at low Galactic latitudes, and the cosmic infrared background dominates at high Galactic latitudes. The signal-to-noise-ratio (SNR) has therefore been adapted for each particular case. In order to increase the completeness and explore possibly interesting new sources at fainter flux density levels, the initial overall reliability goal of the PCCS was reduced to 80\%. The S/N thresholds applied to each frequency channel have been determined, as far as possible, to meet this goal. 

The reliability of the catalogues has been assessed using internal and external validation. At 30, 44, and 70\,GHz, the reliability goal alone would permit S/N thresholds below 4. A secondary goal of minimizing the upward bias on fainter flux densities led to the imposition of an S/N threshold of 4. At higher frequencies, where the confusion caused by the Galactic emission starts to become an issue, the sky has been divided into two zones, one Galactic (52\% of the sky) and one extragalactic (48\% of the sky), using the G45 mask defined in Planck Collaboration XV (2013)~\cite{plxv}. At 100, 143, and 217\,GHz, the S/N threshold needed to achieve the target reliability is determined in the extragalactic zone, but applied uniformly on sky. At 353, 545, and 857\,GHz, the need to control confusion from Galactic cirrus emission led to the adoption of different S/N thresholds in the two zones. This strategy ensures interesting depth and good reliability in the extragalactic zone, but also high reliability in the Galactic zone. The S/N thresholds and additional characteristics of the PCCS are given in Table \ref{tab:table}.

\begin{figure}
\begin{minipage}{1\linewidth}
\centerline{\includegraphics[width=1\linewidth]{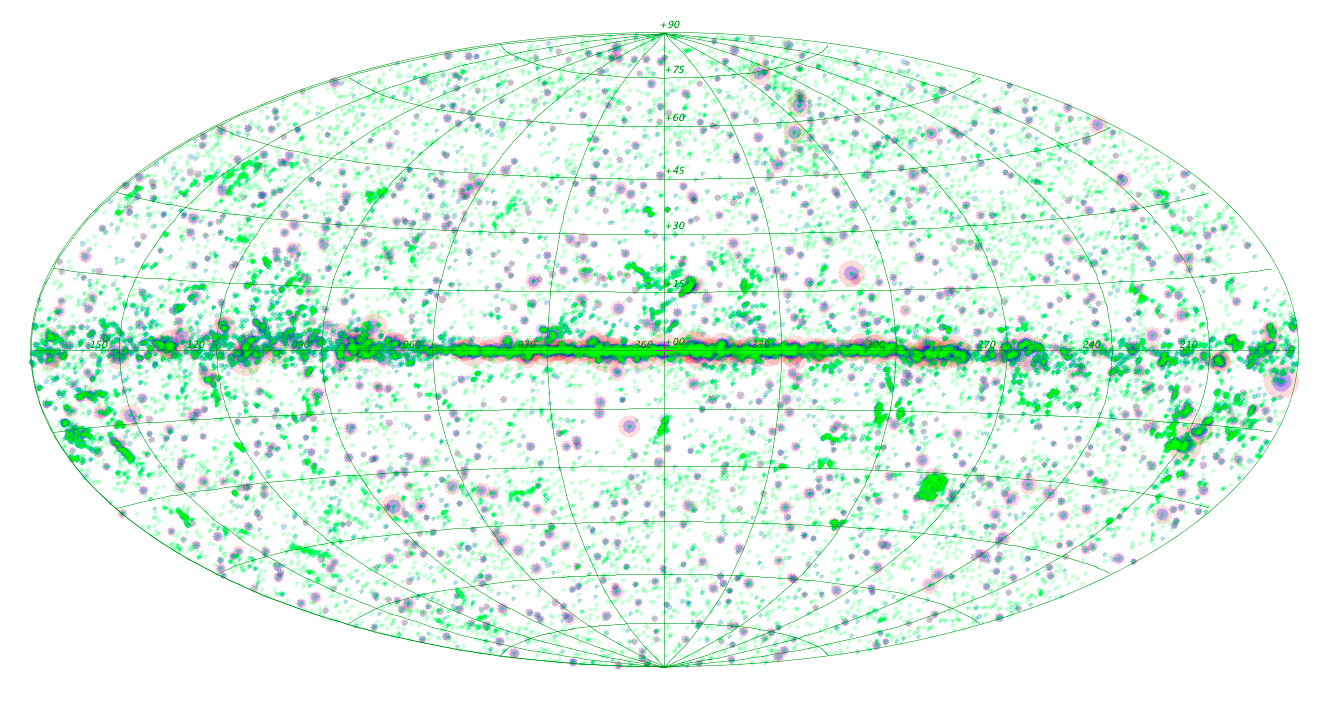}}
\end{minipage}
\caption{Sky distribution of the PCCS sources in three different channels: 30\,GHz (pink circles); 143\,GHz (magenta circles); and 857\,GHz (green circles). The dimension of the circles is related to the brightness of the sources and the beam size of each channel. The figure is a full-sky Aitoff projection with the Galactic equator horizontal; longitude increases to the left with the Galactic centre in the centre of the map.}
\label{fig:plot_pccs}
\end{figure}

\begin{table}
\centering
\caption{PCCS characteristics}
\label{tab:table}
%\begin{tabular}{l@{ \hspace{0.1cm} }c@{ \hspace{0.1cm} }c@{ \hspace{0.1cm} }c@{ \hspace{0.1cm}
%    }c@{ \hspace{0.1cm} }c@{ \hspace{0.8cm} }c@{ \hspace{0.8cm} }c@{ \hspace{0.1cm} }c@{ \hspace{0.1cm} }c}
%
%\verb^\begin{tabular}{|...|}^
\begin{tabular}{|lccccccccc|}
\noalign{\smallskip}\hline
%\noalign{\smallskip}
Channel & 30 & 44 & 70 & 100 & 143 & 217 & 353 & 545 & 857\\\hline\hline
%\noalign{\vskip 3pt\hrule\vskip 5pt}
Freq [GHz] & 28.4 & 44.1 & 70.4 & 100.0 & 143.0 & 217.0 & 353.0 & 545.0 & 857.0 \\
$\lambda$ [$\mu$m] & 10561 & 6807 & 4260 & 3000 & 2098 & 1382 & 850 & 550 & 350 \\
Beam FWHM$^{a}$  [arcmin] & 32.38 & 27.10 & 13.30 & 9.65 & 7.25 & 4.99 & 4.82 & 4.68 & 4.33 \\ \hline
%\noalign{\vskip 10pt}
%\noalign{\textit{ SNR thresholds} }
\textit{SNR thresholds} & &  & &  &  &  & &  & \\
Full sky & 4.0 & 4.0 & 4.0 & 4.6 & 4.7 & 4.8 & \ldots & \ldots & \ldots\\
Extragactic zone$^{b}$ & \ldots & \ldots & \ldots & \ldots & \ldots & \ldots & 4.9 & 4.7 & 4.9\\
Galactic zone$^{b}$ & \ldots & \ldots & \ldots & \ldots & \ldots & \ldots & 6.0 & 7.0 & 7.0\\ \hline
%\noalign{\vskip 10pt}
\textit{Number of sources} & &  & &  &  &  & &  & \\
Full sky & 1256 & 731 & 939 & 3850 & 5675 & 16070 & 13613 & 16933 & 24381 \\
$|b|>30\deg$ & 572 & 258 & 332 & 845 & 1051 & 1901 & 1862 & 3738 & 7536 \\ \hline
%\noalign{\vskip 10pt}
\textit{Flux densities} & &  & &  &  &  & &  & \\
Minimum$^{c}$ [mJy] & 461 & 825 & 566 & 266 & 169 & 149 & 289 & 457 & 658 \\
90\,\% completeness [mJy] & 575 & 1047 & 776 & 300 & 190 & 180 & 330 & 570 & 680 \\
Flux density error [mJy] & 109 & 198 & 149 & 61 & 38 & 35 & 69 & 118 & 166 \\
%\noalign{\vskip 10pt}
Position $^{d}$ [arcmin] & 1.8 & 2.1 & 1.4 & 1.0 & 0.7 & 0.7 & 0.8 & 0.5 & 0.4 \\ \hline
%\noalign{\smallskip}\hline
%\noalign{\smallskip}
\end{tabular}
\begin{tablenotes}
\footnotesize
\item $^\textit{a}$ FEBeCoP band-averaged effective beams. 
\item $^\textit{b}$ The Galactic and extragalactic zones are defined in Planck Collaboration XXVIII (2013)~\cite{pccs}.\par
\item $^\textit{c}$ Minimum flux density of the catalogue at $|b|>30\deg$ after excluding the faintest 10\,\% of sources.\par
\item $^\textit{d}$ Positional uncertainty derived by comparison with the PACO sample (M.~Massardi {\it et al.}~\cite{paco}) up to 353\,GHz and with \textit{Herschel} samples in the other channels.\par
\end{tablenotes}
\end{table}

\begin{figure}
\begin{center}
\begin{minipage}{0.5\linewidth}
\centerline{\includegraphics[width=1\linewidth]{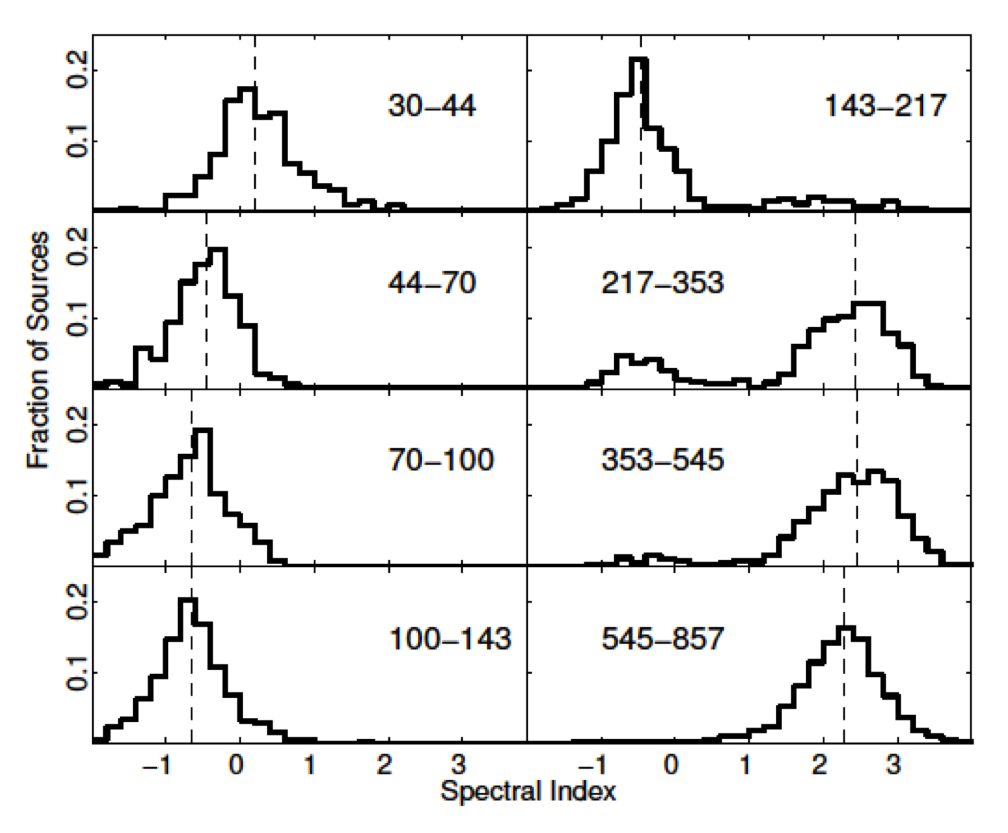}}
\end{minipage}
\caption{Histograms of the spectral indices obtained from the matches between contiguous channels. 545 and 857\,GHz bands are dominated ($> 90 \%$) by dusty galaxies and 30-143GHz bands are dominated ($> 95\%$) by synchrotron sources. The difference between the median values of the spectral indices below 70\,GHz indicate that there is a steepening in blazar spectra; and the high frequency counts ($<217$\,GHz) of extragalactic sources are dominated at the bright end by synchrotron emitters, not dusty galaxies.}
\end{center}
\label{fig:specindex}
\end{figure}

\begin{figure}
\begin{center}
\begin{minipage}{0.5\linewidth}
\centerline{\includegraphics[width=1\linewidth]{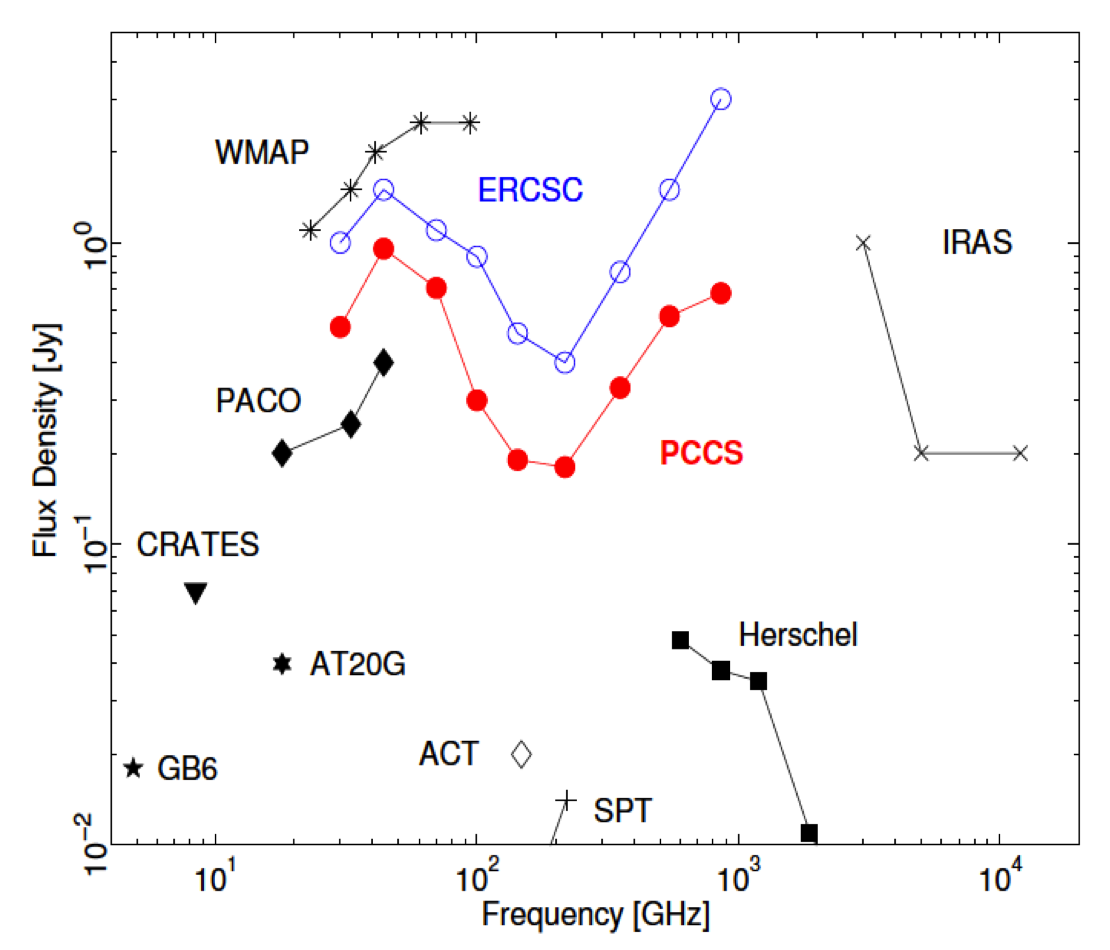}}
\end{minipage}
\caption{Completeness level of PCCS at high Galactic latitude ($|b| > 30$) relative to the ERCSC and other wide area surveys at comparable frequencies. In this figure it can be seen that the sensitivity of the PCCS is better than that of the PCCS and \textit{WMAP}.}
\end{center}
\label{fig:sensit}
\end{figure}

\section{Validation of the Catalogue}
The PCCS contents and the four different flux-density estimates have been validated by simulations (internal validation) and comparison with other observations (external validation). The validation of the non-thermal radio sources was done with a large number of existing catalogues, whereas the validation of thermal sources is mostly done with simulations. Detections identified with known sources have been marked in the catalogues.

The LFI catalogues have been validated using the following catalogues: NEWPS (M. L\'opez-Caniego {\it et al.}~\cite{newps1} and M. Massardi {\it et al.}~\cite{newps2}), AT20G (T. Murphy {\it et al.}~\cite{atca}), CRATES (S.~E.~Healey {\it et al.}~\cite{crates}) and the ERCSC (Planck Early Results VII~\cite{ercsc}). The HFI catalogues have primarily been validated through an internal Monte Carlo quality assessment (QA) process in which artificial sources are injected in both real maps and simulated maps, and to some extent, with data from \textit{Herschel} and other experiments. In the HFI,  the detection is described by the completeness and reliability of the catalogue: completeness is a function of intrinsic flux density, the selection threshold applied to detection (S/N), and location, while reliability is a function only of the detection S/N. The quality of the HFI photometry and astrometry is assessed by direct comparison of detected position and flux density with the known parameters of the artificial sources. An input source is considered to be detected if a detection is made within one beam FWHM of the injected position. 

\section{Photometry}
For each source in the PCCS we have obtained four different measures of the flux density. They are determined by the source detection algorithm (see M. L\'opez-Caniego {\it et al.}~\cite{comp} for further details); aperture photometry; point spread function (PSF) fitting; and Gaussian fitting. Only the first is obtained from the filtered maps, and the other measures are estimated from the full-sky maps at the positions of the sources. The source detection algorithm photometry, the aperture photometry and the PSF fitting use the \textit{Planck} band-average effective beams, calculated with FEBeCoP (Fast Effective Beam Convolution in Pixel Space) as described in S. Mitra {\it et al.}~\cite{febecop}. Notice that only the PSF fitting uses a model of the PSF that depends on the position of the source and the scan pattern. When available, we have validated the photometry of the PCCS with external data, ground-based follow-ups and with \textit{Herschel} data.

\section{Statistical properties of the PCCS}
Figure~\ref{fig:specindex} shows histograms of the spectral indices obtained from the matches between contiguous channels. As expected, the high-frequency channels (545 and 857\,GHz) are dominated ($>90\,\%$) by dusty galaxies and the low-frequency channels are dominated ($>95\,\%$) by synchrotron sources. In addition, this Figure shows that the PCCS confirms  two striking results from the ERCSC:
%obtained making use of the ERCSC are evident in Figure~\ref{fig:specindex}:
 the difference between the median values of the spectral indices below 70\,GHz indicates that there is a significant steepening in blazar spectra, as demonstrated in Planck Early Results XIII~\cite{planck2011-6.1}; and the high-frequency counts (at least for frequencies $\leq217$\,GHz) of extragalactic sources are dominated at the bright end by synchrotron emitters, not dusty galaxies (Planck Intermediate Results VII~\cite{planck2012-VII}). The deeper completeness levels and, as a consequence, the higher number of sources compared with  the ERCSC, will allow the extension of previous studies to more sources and to fainter flux densities (see Figure~\ref{fig:sensit}). 

\section{Conclusions}
The PCCS contains sources extracted from the \textit{Planck} nominal mission data in each of its nine frequency bands. By construction its reliability is $> 80\%$ and a special effort was made to use simple selection procedures in order to facilitate statistical analyses. With a common detection method for all the channels and four photometries, spectral analysis can be performed safely. The deeper completeness levels and, as a consequence, the higher number of sources compared with its predecessor the ERCSC, will allow the extension of previous studies to more sources and to fainter flux densities. The PCCS is the natural evolution of the ERCSC, but both lack polarization and multi-frequency information. Future releases of the PCCS will take advantage of the increased sensibility of the \textit{Planck} full mission data, with more sources and polarization information that have not been included in the 2013 data release. 

\section*{Acknowledgements} 
The development of \textit{Planck} has been supported by: ESA; CNES and CNRS/INSU-IN2P3-INP (France); ASI, CNR, and INAF (Italy); NASA and DoE (USA); STFC and UKSA (UK); CSIC, MICINN, JA and RES (Spain); Tekes, AoF and CSC (Finland); DLR and MPG (Germany); CSA (Canada); DTU Space (Denmark); SER/SSO (Switzerland); RCN (Norway); SFI (Ireland); FCT/MCTES (Portugal); and PRACE (EU). A description of the \Planck\ Collaboration and a list of its members, including the technical or scientific activities in which they have been involved, can be found at \url{http://www.sciops.esa.int/index.php?project=planck&page=Planck_Collaboration}.

\end{document}